# An electric circuital analysis of laboratory plasma sheath fluctuations and propagations

Subham Dutta and Pralay Kumar Karmakar[*]
Department of Physics, Tezpur University, Napaam-784028, Tezpur, Sonitpur, Assam, India
[*]E-mail: pkk@tezu.ernet.in

**ABSTRACT**

The effective inductive ($L$), capacitive ($C$), and resistive ($R$) behavior of a plasma sheath in a conjoint coupled form is well familiar among plasma physics communities. A dynamic sheath instability in laboratory plasmas is systematically modelled herein as an electrical series-resonance *LCR* circuit of the above kind. It theoretically yields experimentally observed findings on coexistent plasma sheath oscillation, electric current perturbation, and subsequent plasma sheath waves (PSWs). The plasma current in the *LCR* circuit formalism is allowed to undergo linear (small-scale) spatiotemporal perturbation about its homogeneous equilibrium state. The oscillating sheath triggers ion-acoustic wave excitation in the bulk plasma through sheath-induced energy transfer processes. The obtained results could be applicable mainly in understanding electromagnetic communication antennas, ion energy modulation processes, diverse plasma probe diagnostics, etc.

**Introduction**

It is well known that plasma never comes into direct contact with the confining boundary walls. A nonlinear nonneutral space charge layer, termed as 'plasma sheath' or 'Debye sheath' having a width comparable to the Debye scale size, is bound to be formed near the vicinity of the plasma-confining wall. Such a sheath formation is a ubiquitous phenomenon in any plasma medium at the junction of these bulk plasma and its confining chamber walls. Such a nonlinear charge layer gets developed also around any electrodes, or similar objects introduced in the plasma medium. It shows several atypical characteristics in terms of the electric current versus voltage relationship [1], plasma stability [2], various instability phenomena [3], and so forth.

Apart from the above, a laboratory plasma sheath often also resembles sheaths formed in the astroplasmic scenarios, such as the Gravito-electrostatic sheath (GES) in the Sun [4], magnetospheric sheath in the magnetars [5], electromagnetic sheath in the astronomic antennas [6], etc. Although these sheaths vary in terms of their properties and formation mechanism, they also resemble in some characteristic features. Therefore, studying plasma sheaths and their evolutions using various model formalisms is highly important from both the fundamental and applied perspective. These model formalisms help in exploring the plasma medium, analyzing various plasmic phenomena, studying waves and instabilities under different relevant circumstances. The suitability of the model consideration depends on the typical orders of the involved plasma parameters which vary with the environments of plasma existence.

In this regard, it may be added that the discussed plasma models are comprised of the single particle description [7], the kinetic description [8], the fluid description [9], the hybrid-kinetic fluid description [10], the gyrokinetic description [11], and so forth [12]. These plasma model formalisms are applied in respective circumstances due to their corresponding feasible results therein.

The formalisms discussed above are vividly known and well-explored across the plasma physics community. However, in this work, motivated by different formalisms as of now as stated above, we study laboratory plasma sheath using an unconventional series *LCR* circuit model [13,14] through a local linear normal perturbation formalism. It helps in inculcating the formation of plasma sheath wave (PSW) [15] and possible energy transfer to the bulk plasma through plasma ion-acoustic wave (IAW) formations [16], electron wave formations in the *LCR* circuital model perspective.

A circuital plasma model has also been developed before [17]. However, there are some differences between the earlier models with the current one described in this presented analysis. In the earlier model studies, the sheath is considered only as a capacitive element, ignoring its electrical resistive and inductive properties. However, in this modified *LCR* circuital model, the sheath shows simultaneously inductive, capacitive, and resistive properties in series against the different types of mixed (non-series) combinations of the circuit elements considered previously in the literature [17].

The modified *LCR* circuital model formalism of plasma sheath can be justified with its simultaneous inductive, capacitive, and resistive behavior during various plasmic operations. Inductive behavior is known to be an outcome of undulating linear temporal electric field fluctuations across the dynamic sheath. This temporal electric field fluctuation triggers directional variation of the ion current and generates the spatially varying magnetic field. The energy stored in the form of a magnetic field yields the inductive nature of the sheath circuit. The inductor resists any change in the current which boosts the energy storage in the inductor. However, the noticed negative polarity of the inductance ($L$) herein denotes the induced *emf* supporting the primary cause of current variation across the sheath. It implies that an increase or decrease in the current across the sheath continues until a current reversal occurs therein [18].

The capacitive behavior originates from the bipolar nature of the plasma sheath which resembles an opposite polarity parallel plate capacitor with the plate separation comparable to the super-Debye scale. During the sheath formation, the electron accumulation on the chamber wall and electron decumulation (dispersal) at the adjacent region (sheath) form a bipolar region of charge distribution, both having opposite polarities. This stiff difference in electrostatic potential within such a small spatial distance ($\sim \lambda_D$) generates a strong electric field amidst the two sides, viz., the chamber wall (negative) and sheath-plasma junction region (less negative). This separation is equivalent to the capacitive plate separation. These two regions collectively replicate that of a parallel plate capacitor as an electric energy storage device. Furthermore, the intermediate region aids in strengthening the electric field, acting as an intrinsic dielectric material, as arranged externally in real capacitors used in engineering and technology [18].

Apart from the above, the resistive behavior of the plasma sheath is a consequence of the potential barrier developed across the two opposite polarity sides of the sheath. The only charges with or above the required threshold velocity and kinetic energy manage to get across the sheath potential barrier. This resistive behavior differs from an ordinary Ohmic resistor in terms of potential versus the current relationship. The *emf* generated across the sheath is equivalent to the acquired negative potential on the plasma boundary. The potential gradient of the sheath induces the ions to move across it. This changes the spatial boundary of the sheath edge, and thus aids in the nonzero temporal variation of the effective *emf* across the sheath [18].

Besides, the PSWs studied herein are a special kind of wave in the plasma medium which originate and propagate across the sheath region. The strong electric field within a sheath and non-equilibrium conditions in terms of different plasma parameters influence the PSWs therein. The PSW may get excited upon the application of adequate perturbation in the sheath region. Depending on their formation mechanisms, the PSWs may be primarily categorized as electrostatic PSWs (occurs through electrostatic potential fluctuations) [19], electromagnetic PSWs (occurs through electromagnetic wave interaction with sheaths, such as polaritons) [20], and surface waves (occurs through the coupling between PSWs and surface waves) [21].

Some of the key features of PSWs also seem to have considerable applied values. The PSWs can transfer energy from the sheath to the plasma boundary surface leading to sputtering and material heating-like phenomena [22]. In radio-frequency (RF) plasmas, the PSWs can influence constituent ion energy distribution and are relevant for analyzing etching and deposition processes [23,24]. The PSWs can play a role in systems with oscillating boundaries, such as antennas or tethers [25]. These PSWs are relevant from both the fundamental plasma physics background as well as applied fields across the domains. Such waves provide physical insights about the plasma sheath properties in terms of the electrostatic potential distribution, sheath thickness, ion dynamics including implantation, and so forth [22,26-28].

The PSWs evolving on the Debye spatiotemporal scales can also excite IAWs in the bulk plasma system. The development of the IAW from the PSWs begins with the perturbation in the sheath electric field, thereby making the fluctuation. As a consequence, the electric charge densities in the sheath start oscillating simultaneously. Such perturbations in the sheath region can propagate across the sheath-plasma interface and induce similar parametric perturbations in the adjacent plasma accordingly. The constitutive ions in the electron-deficient sheath are accelerated by the perturbed electric field, thereby causing oscillations in both the ion velocity and ion number density. These oscillations trigger the excitation of the IAWs in the bulk plasma.

The IAWs are the low-frequency longitudinal waves in plasmas, analogous to the usual sound waves, excitable in neutral gaseous media. The IAWs originate due to the pressure gradients in the electron (thermal species) population density generating thermal pressure force, while the ions (inertial species) provide inertia. Thus, the restoring force of the oscillating ions originates from the charge separation and resultant electrostatic polarization force [29].

The transition region between the sheath and quasineutral plasma acts as a coupling point between these two regions. The factors that contribute to the coupling are adequate sheath thickness, ion temperature, nonlinear effects, and external drivers. The usual conditions that must be fulfilled for the IAW excitation from the PSWs are frequency matching (between the excited PSW and excitable IAW), smooth interlayer matching conditions, etc. [16]. In other words, the PSWs and IAWs have been studied theoretically by using the discussed non-circuital model formalisms with considerably corroborating results as reported in the past [15,16]. After a theoretical *LCR* circuital model development, we conjecture herein for the first time an inductor ($L$), capacitor ($C$), and resistor ($R$) model of the plasma sheath to study PSWs and energy propagation from the oscillating sheath to the bulk plasma through IAWs. This model adequately accounts for the efficient energy transfer from the sheath to the bulk plasma.

One of the possible ways of exciting the PSWs and IAWs is through using a capacitively coupled plasma (CCP) discharge system. Along with the PSWs and IAWs, another wave which gets excited in special circumstances in the CCPs, is the electron plasma waves (also termed as the electron acoustic waves or Langmuir waves). An RF voltage is applied across two installed electrodes in the plasma system [30]. It yields a time-varying electric field penetrating the plasma and results in plasma sheath oscillations. During the RF oscillations, the plasma-sheath interface undergoes periodic expansion and contraction. The time-varying boundaries can launch electron beams (also, causes collisionless electron heating) [31-33], inducing electron plasma waves due to charge imbalance and perturbations in the CCP system. The perturbed electrons from their respective equilibrium positions oscillate around that point due to the restoring electric fields. It leads to the excitation of electron plasma waves [34]. It may further be added that these waves are obviously longitudinal in nature, expressed in the form of spatiotemporal electron density variations [35,36]. The electron motion in the system can trigger nonlinear kinetics, wave-particle interactions, and excitation of kinetic electron waves. The choice of waveforms is critical for density driven waves [37].

This article, apart from the introduction part as already presented in section 1, is organizationally structured as follows. Section 2 elaborately discusses the circuital model and associated longitudinal wave excitations. Section 3 gives the mathematical model in terms of the governing differential equation and solves it with linear small-scale perturbation formalism. Then, section 4 offers the solutions of the governing equation with their respective self-illustrative parametric profiles. Lastly, section 5 conclusively summarizes the main implications and applications of the circuital model analysis in a broad applied horizon.

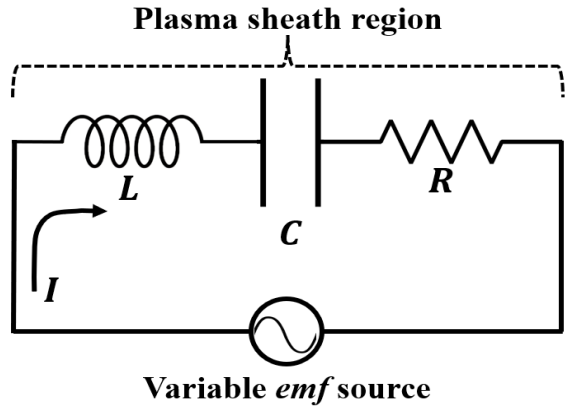

**Figure 1**. A schematic diagram manifesting the *LCR* circuital behavior of a plasma sheath with all the relevant circuital components in series combination.

**Mathematical modelling**
The *LCR* circuit model has been proposed before for dynamic transient sheaths with reasonable results [13,16]. We apply the Kirchhoff voltage law (KVL) for the series *LCR* combination and model the sheath circuit theoretically by adding the individual potential drops across the circuit components for studying the transient plasma sheath. Considering the series *LCR* components (Fig. 1), the governing voltage equation for the above circuital combination is cast as

$$V_L + V_R + V_C = e. \tag{1}$$

Here, the terms $V_L$, $V_R$, $V_C$, and $e$ denote the voltage drops across the inductor, resistor, capacitor, and source *emf* (electric potential difference across the sheath), respectively. Using their known expressions in terms of individual circuital components and temporal current variations of different orders, viz., $V_L = L\partial_t I$, $V_R = RI$, and $V_C = Q/C$, we get

$$L\partial_t I + RI + \frac{Q}{C} = e. \tag{2}$$

Here, $L$, $C$, $R$, and $Q$ denote the sheath inductance, capacitance, resistance, and charge stored within the sheath capacitor, respectively. Whereas $I$ denotes the current through the sheath (also termed as ion implantation current) and $\partial_t (= \partial/\partial t)$ is the first-order partial derivative with respect to time. Moreover, since we notice a spatiotemporal variation of the sheath width ($x$) resulting in *emf*-variation, therefore, we differentiate Eq. (2) with respect to time to account for the temporally changing *emf* across the sheath. Using $\partial_t Q = I$, the second-order differential form of Eq. (2) is cast as follows

$$L\partial_t^2 I + R\partial_t I + IC^{-1} = \partial_t e. \tag{3}$$

Here, $\partial_t^2 (= \partial^2/\partial t^2)$ is second-order partial derivative with respect to time. Moreover, for a sheath with effective negative inductance ($L < 0$) condition [13], Eq. (3) is modified as

$$L\partial_t^2 I - R\partial_t I - IC^{-1} = -\partial_t e. \tag{4}$$

Further, the sheath has a transient nature with its oscillating edge towards and away from the plasma. The moving edge changes the temporal variation of the sheath *emf* through its associated spatial width variation. Therefore, the term designating temporal *emf* variation on the RHS of Eq. (4), i.e., $-\partial_t e$ is modified as $(-\partial e/\partial x)(\partial x/\partial t)$ or $(-\partial_x e)(\partial_t x)$. Here, $\partial_x e$ denotes the spatial variation of *emf* and $\partial_t x$ denotes temporal variation of sheath width. Incorporating these rearrangements, Eq. (4) in a modified form now reads as

$$L\partial_t^2 I - R\partial_t I - IC^{-1} = (-\partial_x e)(\partial_t x). \tag{5}$$

Further, in an *LCR* circuit, the *emf*, $e$ can be expressed in terms of $I$ and $Z$ as $e = IZ$. Here, $Z$ is impedance expressed as $Z (= \sqrt{R^2 + (X_L - X_C)^2})$ through the circuit parameters $R$, $X_L$, and $X_C$ denoting resistance, inductive reactance, and capacitive reactance, respectively. Since these individual circuit parameters ($R$, $X_L$, and $X_C$) are assumed spatiotemporally invariant, therefore the impedance $Z$ is also constant. Using the following in Eq. (5), one gets

$$L\partial_t^2 I - R\partial_t I - IC^{-1} = (-Z\partial_x I)(\partial_t x). \tag{6}$$

We replace the $\partial_t x$-term in the RHS of Eq. (6) with the sound phase speed in the plasma, $c_s$. This can be justified by the temporal sheath width variation in the order of $c_s$. Using Eq. (2) and Eq. (6), we have

$$L\partial_t^2 I - R\partial_t I - IC^{-1} = -c_s Z \partial_x I. \tag{7}$$

Through a multiplication of $C$ on both sides of Eq. (7), we have

$$LC\partial_t^2 I - CR\partial_t I - I = -c_s CZ \partial_x I. \tag{8}$$

It is assumed that all parameters, except $I$ in Eq. (8) are spatiotemporally invariant. For studying the sheath dynamics, we perturb Eq. (8) through the substitution of $I$. The sheath current parameter $I$ is assumed to undergo linear small-scale perturbation ($I_1$) about its equilibrium value ($I_o (\gg I_1)$) in the following way

$$I = I_o + I_1 = I_o + I_{10} \exp\{-i(\omega t - kx)\}. \tag{9}$$

Using this common perturbation scheme of $\partial_x \equiv ik$, $\partial_t \equiv -i\omega$, and $\partial_t^2 \equiv -\omega^2$ [38] from Eq. (9) in Eq. (8) and removing the common parameter $I_{10}$ from both sides, we have

$$(LC)\omega^2 + (-iCR)\omega + (1 - ic_s CZk) = 0. \tag{10}$$

Here, Eq. (10) may be solved by comparing it to the quadratic equation of the form $ax^2 + bx + c = 0$. The customary solutions of quadratic equations of such kinds are given by $x = \left(-b \pm \sqrt{b^2 - 4ac}\right)/2a$. Applying the same in Eq. (10), we have its solution given as

$$\omega = (2LC)^{-1}\left(iCR \pm \sqrt{-C^2R^2 - 4LC(1 - ic_s CZk)}\right). \tag{11}$$

The term with square root in Eq. (11) can be simplified by comparing it to $a + ib$ as

$$\sqrt{-C^2R^2 - 4LC(1 - ic_s CZk)} = a + ib. \tag{12}$$

The expressions for $a$ and $b$ can be found by squaring Eq. (12) and comparing the real and imaginary terms on both sides. After some simple rearrangements, one gets

$$a = \pm 2c_s LC^2 Z \left[\sqrt{0.5}\left\{(C^2R^2 + 4LC) + \sqrt{(C^2R^2 + 4LC)^2 + 16c_s^2 L^2 C^4 Z^2 k^2}\right\}^{-\frac{1}{2}}\right] k, \tag{13}$$

$$b = \pm\sqrt{0.5}\left[(C^2R^2 + 4LC) + \sqrt{(C^2R^2 + 4LC)^2 + 16c_s^2 L^2 C^4 Z^2 k^2}\right]^{\frac{1}{2}}. \tag{14}$$

Substituting the values of $a$ and $b$ in Eq. (12) and then in Eq. (11), we have the complete solution of $\omega$ in terms of its real component ($\omega_r$) and imaginary component ($\omega_i$) given as

$$\omega = (2LC)^{-1}\left[\left\{iCR \pm i\sqrt{0.5}\left\{(C^2R^2 + 4LC) + \sqrt{(C^2R^2 + 4LC)^2 + 16c_s^2 L^2 C^4 Z^2 k^2}\right\}^{\frac{1}{2}}\right\} \pm \right.$$

$$\left. \pm 2c_s LC^2 Z \left\{\sqrt{0.5}\left\{(C^2R^2 + 4LC) + \sqrt{(C^2R^2 + 4LC)^2 + 16c_s^2 L^2 C^4 Z^2 k^2}\right\}\right\}^{-\frac{1}{2}} k\right]. \tag{15}$$

We now use a complex frequency, $\omega = 1\omega_r + i\omega_i$ in Eq. (15), to characterize the overall PSW mode in the presence of available free energy (plasma currents) across the sheath in the considered plasma model system. Here, '1' is the real unit and $\omega_r$ is the real frequency part, depicting the normal and regular instability propagatory features. Then, '$i$' is the imaginary unit and $\omega_i$ is the imaginary frequency part characterizing the abnormal and irregular instability growth features. It enables us to split Eq. (15) in terms of the frequency conjugations as

$$\omega_r = \pm\sqrt{2}c_s CZ\big[(C^2R^2 + 4LC) + \sqrt{(C^2R^2 + 4LC)^2 + 16c_s^2 L^2 C^4 Z^2 k^2}\big]^{-\frac{1}{2}} k, \qquad (16)$$

$$\omega_i = (2LC)^{-1}\left[\pm\sqrt{0.5}\big\{(C^2R^2 + 4LC) + \sqrt{(C^2R^2 + 4LC)^2 + 16c_s^2 L^2 C^4 Z^2 k^2}\big\}^{\frac{1}{2}} + CR\right]. \qquad (17)$$

We now use $D$ to denote the algebraic discriminant of Eq. (10) in a modified form as

$$D = \sqrt{0.5}\big[(C^2R^2 + 4LC) + \sqrt{(C^2R^2 + 4LC)^2 + 16c_s^2 L^2 C^4 Z^2 k^2}\big]^{\frac{1}{2}}. \qquad (18)$$

Consequently, Eq. (17) becomes

$$\omega_i = (2LC)^{-1}[\pm D + CR]. \qquad (19)$$

The following comments on the above instability formalism may be worth mentioning. Here, it is repeated that $\omega_r$ denotes the propagative ($\omega_r > 0$), standing ($\omega_r = 0$), and evanescent ($\omega_r < 0$) nature of the PSW (as in Eq. (16)). Then, $\omega_i$ denotes the growth ($\omega_i > 0$, instability), steady state ($\omega_i = 0$, oscillatory), and decay ($\omega_i < 0$, stability) of the PSW (as in Eq. (17)). It is evident from Eq. (19) that there is a possibility of having $\omega_i > 0$, since $D > RC$ at some situations. It denotes the excitation of plasma acoustic instabilities due to the electric current perturbation resulting from the sheath oscillations. The free energy source for this instability excitation exists in the electric field generated due to the deviation of the system from quasineutrality. The resultant stiff population density gradient promotes the spatial propagation of the instability pulses from the sheath to the bulk plasma.

We derive the resonance condition at which different plasma waves get excited due to the plasma sheath interaction in terms of the natural electric current oscillation frequency ($\omega_c$) across the $LCR$ circuit equivalent sheath. It is assumed that the oscillating sheath produces PSW (due to its propagative nature) and its frequency ($\omega_r$) equals the frequency of the plasma acoustic waves propagating from the sheath towards the bulk plasma. Since the terms $L$, $C$, and $R$ are invariant, the impedance ($Z$) deals with only one variable, i.e., $\omega_c$. The term $\omega_c$ governs the inductive and capacitive reactance of the plasma sheath under consideration as

$$Z = \sqrt{R^2 + (\omega_c L - (\omega_c C)^{-1})^2}. \qquad (20)$$

We derive the condition of plasma acoustic wave excitation by expressing Eq. (16) in terms of $\omega_c$. Replacing the magnitude of $\omega_r$ with any plasma acoustic wave frequency and substituting the values of other relevant parameters, the order of $\omega_c$ can be evaluated from Eq. (16). However, before evaluating $\omega_c$, we figure out the approximate values of the circuit parameters $L$, $C$, and $R$. This expedites the simplification of expressions by ignoring the small-order redundant terms. The expressions derived for such parameters [14] are given as

$$L = -(c_s \tau_{io}^3 \lambda^2)(40\epsilon_o)^{-1}, \qquad (21)$$

$$C = \epsilon_o(1 + 0.17\lambda^2) X_O^{-1}, \qquad (22)$$

$$R = c_s \lambda^2 \tau_{io}^2 (12\epsilon_o)^{-1}. \tag{23}$$

Here, $\tau_{io}$ is the ion transit time through the sheath. $\lambda \sim \omega_{pi}\tau_{io}$, typifies the characteristic sheath width and $\omega_{pi}$ is the ion plasma oscillation frequency. The equilibrium sheath width is denoted by $X_o = \{1 + 0.17\lambda^2\}c_s\tau_{io}$ and permittivity of free space, $\epsilon_o = 8.85 \times 10^{-12}$ m$^{-3}$ kg$^{-1}$ s$^4$ A$^2$. The independent terms in Eqs. (21)-(23) can be further expressed as $c_s = (k_B T_e m_i^{-1})^{1/2}$, $\tau_{io} \approx \lambda_D c_s^{-1}$, where $\lambda_D$ typifies the sheath width and $c_s$ is the ion sound phase speed. We use the experimentally relevant values for the various terms, such as thermal energy $k_B T_e \sim 2$ eV; equilibrium number density, $n_o = 10^{14}$ m$^{-3}$; and ionic mass, $m_i = 1.67 \times 10^{-27}$ kg (proton mass exactly for the exemplary hydrogen plasma). Using the above term values and expressions, the typical values of the sheath inductance, capacitance, and resistance are estimated as $L = -1.36 \times 10^{-7}$ H, $C = 8.84 \times 10^{-8}$ F, and $R = 62.5$ Ω, respectively. These parametric values yield qualitative plots of the dependent parameters and aid in examining the behavior of the oscillating sheath.

We simplify the denominator of Eq. (16) by taking the common of the term $(C^2R^2 + 4LC)^2$ from the terms inside the square root as $(C^2R^2 + 4LC)\{1 + 16c_s^2 L^2 C^4 Z^2 (C^2R^2 + 4LC)^{-2} k^2\}^{1/2}$. Substituting the evaluated approximate values of $L, C$, and $R$ in this expression, we figure out $16c_s^2 L^2 C^4 Z^2 (C^2R^2 + 4LC)^{-2} \ll 1$. Therefore, using the approximation of $(1 + x)^{1/2} \approx 1 + x/2$ for $x \ll 1$, we have $\sqrt{0.5}\{2(C^2R^2 + 4LC) + 8c_s^2 L^2 C^4 Z^2 (C^2R^2 + 4LC)^{-1} k^2\}$. Using these modifications in Eq. (16), we have

$$\omega_r = \pm\sqrt{2}c_s(C^2R^2 + 4LC)CZ\{2(C^2R^2 + 4LC)^2 + 8c_s^2 L^2 C^4 Z^2 k^2\}^{-1/2} k. \tag{24}$$

We consider the positive case of $\omega_r$ as it points towards a propagatory nature of sheath oscillation, resulting in the formation of an PSW and efficient energy flow to the bulk plasma. Rearranging Eq. (24) after squaring both sides, we have

$$Z^2 k^2 = \{(C^2R^2 + 4LC)(c_s C)^{-2}\omega_r^2 + 4L^2 C^2 Z^2 \omega_r^2 (C^2R^2 + 4LC)^{-1} k^2\}, \tag{25}$$

Rearranging it further, we have

$$(Zk)^2\{1 - 4L^2 C^2 \omega_r^2 (C^2R^2 + 4LC)^{-1}\} = (c_s C)^{-2}\omega_r^2 (C^2R^2 + 4LC), \tag{26}$$

Simplifying it more, we get

$$(Zk)^2 = (C^2R^2 + 4LC)^2 \omega_r^2 \{(C^2R^2 + 4LC) - 4L^2 C^2 \omega_r^2\}^{-1}(c_s C)^{-2}. \tag{27}$$

Using $Z = \sqrt{R^2 + (\omega_c L - (\omega_c C)^{-1})^2}$ in Eq. (27), we have

$$\{R^2 + (\omega_c L - (\omega_c C)^{-1})^2\}k^2 = \omega_r^2 (C^2R^2 + 4LC)^2\{(C^2R^2 + 4LC) - 4L^2 C^2 \omega_r^2\}^{-1}(c_s C)^{-2}. \tag{28}$$

Multiplying both sides with $\omega_c^2 C^2$, we have

$$\{\omega_c^4 L^2 C^2 + \omega_c^2 C^2 R^2 - 2LC\omega_c^2 + 1\}k^2 = (\omega_c\omega_r)^2 c_s^{-2}(C^2R^2 + 4LC)^2\{(C^2R^2 + 4LC) - 4L^2C^2\omega_r^2\}^{-1}. \tag{29}$$

Using the parametric values of $L$, $C$, and $R$, we find $(C^2R^2 + 4LC) \ll 4L^2C^2\omega_r^2$, therefore we use $(C^2R^2 + 4LC) - 4L^2C^2\omega_r^2 \approx -4L^2C^2\omega_r^2$. Applying the same in Eq. (29), we have

$$\omega_c^4 L^2 C^2 k^2 + \omega_c^2\{C^2R^2k^2 - 2LCk^2 + (C^2R^2 + 4LC)^2(4c_s^2L^2C^2)^{-1}\} + k^2 = 0. \tag{30}$$

Since, $CR(\sim 10^{-7}) \gg LC(\sim 10^{-14})$, hence we use some approximations, such as $C^2R^2k^2 - 2LCk^2 \approx (CR)^2 k^2$ and $C^2R^2 + 4LC \approx C^2R^2$ in the solution of Eq. (30) given as

$$\omega_c = \pm(\sqrt{2}LC)^{-1}\Big[-\{C^2R^2k^2 + C^2R^4(4c_s^2L^2)^{-1}\} \pm \sqrt{\{C^2R^2k^2 + C^2R^4(4c_s^2L^2)^{-1}\}^2 - 4(LC)^2k^4}\,\Big]^{1/2}k^{-1}. \tag{31}$$

Since, $\{C^2R^2k^2 + C^2R^4(4c_s^2L^2)^{-1}\}^2 \gg 4(LC)^2k^4$, for any values of the involved parameters, therefore using the approximation $(1 + x)^{1/2} \approx 1 + x/2$ in Eq. (31), we have

$$\omega_c = \pm(\sqrt{2}LC)^{-1}\Big[-\{C^2R^2k^2 + C^2R^4(4c_s^2L^2)^{-1}\} \pm \{\{C^2R^2k^2 + C^2R^4(4c_s^2L^2)^{-1}\} - 2(LC)^2\{C^2R^2k^2 + C^2R^4(4c_s^2L^2)^{-1}\}^{-1}k^4\}\Big]^{1/2}k^{-1}. \tag{32}$$

The Eq. (32) has two physical simplifications, presented distinctly in Eqs. (33)-(34) as

$$\omega_c = \pm(\sqrt{2}LC)^{-1}[-2(LC)^2\{(CR)^2k^2 + C^2R^4(4c_s^2L^2)^{-1}\}^{-1}k^4]^{1/2}k^{-1}, \tag{33}$$

$$\omega_c = \pm(\sqrt{2}LC)^{-1}[2(LC)^2\{(CR)^2k^2 + C^2R^4(4c_s^2L^2)^{-1}\}^{-1} - 2\{(CR)^2k^2 + C^2R^4(4c_s^2L^2)^{-1}\}]^{1/2}k^{-1}. \tag{34}$$

We now consider Eq. (34) for our further analysis as it appears relevant simultaneously for higher and lower $k$-magnitudes, proving its generality over the special case of Eq. (33), where only higher $k$-magnitudes yield pertinent $\omega_c$-values. Moreover, Eq. (34) can yield both real and imaginary values of $\omega_c$, with both positive and negative polarities. Using estimated parametric values along with $k \sim 10^4$ rad m$^{-1}$ in Eq. (34), we estimate the $\omega_c$-magnitude as

$$\omega_c \approx 0.5i \times 10^9 \text{ rad s}^{-1}. \tag{35}$$

Now, for $k \sim 10$ rad m$^{-1}$ instead of $k \sim 10^4$ rad m$^{-1}$, but with similar other parametric values used in Eq. (31), the $\omega_c$-magnitude is estimated as

$$\omega_c \approx -7.66i \times 10^{11} \text{ rad s}^{-1}. \tag{36}$$

The quantitative $\omega_c$-values estimated in Eq. (35) and Eq. (36) are imaginary in nature for both smaller and larger $k$-magnitudes.

The imaginary $\omega_c$-nature denotes the overdamping of the natural sheath current oscillations across the sheath. The oscillation frequencies [$\nu_c = \omega_c(2\pi)^{-1}$] at two assumed $k$-values, viz., $k = 10^4$ rad s$^{-1}$ and $k = 10$ rad s$^{-1}$ are $\nu_c = 7.96i \times 10^7$ s$^{-1}$ (from Eq. (35)), i.e., $|\nu_c| = 7.96 \times 10^7$ Hz and $\nu_c = -1.22i \times 10^{11}$ s$^{-1}$ (from Eq. (36)), i.e., $\nu_c = 1.22 \times 10^{11}$ Hz, respectively. These frequencies at respective $k$-values denote the rate at which the sheath current oscillates. A higher $\nu_c$-magnitude denotes greater sheath current oscillation frequency and thus sheath potential fluctuation due to spatiotemporal sheath width variation. An overdamping sheath oscillation (with imaginary $\omega_c$) with higher and lower magnitudes denote higher and lower rates of energy transfer from the sheath to the bulk plasma, respectively. Besides, the conditions which govern the characteristics of the natural current oscillations in an $LCR$ circuit in different cases are cast as

$$R^2 < 4LC^{-1} \text{ (underdamped);} \tag{37}$$

$$R^2 > 4LC^{-1} \text{ (overdamped);} \tag{38}$$

$$R^2 = 4LC^{-1} \text{ (critically damped).} \tag{39}$$

It is quite evident from the estimated values of $L$, $C$, and $R$ that the condition of overdamping, viz., $R^2 > 4LC^{-1}$ (Eq. (38)), is also well fulfilled, as already seen via Eqs. (35)-(36). Therefore, it is once again shown that the sheath current oscillations or PSWs undergo gradual damping after being excited in the plasma system. This damping sheath oscillation energy is converted into exciting the IAW therein. An imaginary $\omega_c > 0$ indicates an instability excitation with energy transfer from the bulk plasma to the oscillating sheath. In contrast, an imaginary $\omega_c < 0$ indicates instability excitation with the energy transfer from the oscillating sheath to the bulk plasma. The first condition reduces the PSW amplitude and increases the IAW amplitude. However, the second condition does exactly the opposite of the amplitudes of these two waves. The free energy source lies in the fluctuating sheath width and resultant population density inhomogeneities. The outcome of the instability is theoretically manifested in the form of temporally growing (without saturation) current perturbations in the sheath $LCR$ circuit. In reality, the perturbation is bound to saturate due to the involved intrinsic nonlinearities.

**Results and discussions**
An $LCR$ circuital model of plasma sheath is constructed herein to study the energy transfer mechanism from the sheath to the bulk plasma through the excited PSWs and plasma acoustic waves. This $LCR$ circuital sheath model has been established successfully earlier through published literature [13,14,16]. The simultaneous inductive ($L$), capacitive ($C$), and resistive ($R$) behaviors of the plasma sheath is a collective outcome of the intrinsic electric field, corresponding potential barrier, and their fluctuations across the sheath width. The values of $L$, $C$, and $R$ are qualitatively estimated from their derived expressions in the literature [14] and using typical laboratory-scale parametric values of the independent parameters involved.

The application of the $LCR$ circuital sheath model yields a nonhomogeneous second-order linear differential equation (DE) with constant coefficients (Eq. (5)). We further apply a small-scale linear perturbation formalism on the dependent ion implantation current parameter: $I_1(x,t) = I_{1o} \exp\{-i(\omega t - kx)\}$ [38]. The current through the sheath undergoes small-scale perturbation with frequency $\omega$ and within a spatial distance of $x$. The nature of the derived $\omega$-parameter shows the characteristics of the PSWs excited.

The application of perturbation formalism reduces the DE into quadratic equation (QE) with variable $\omega$ (Eq. (7)). The QE is solved analytically using a set of justifiable approximations without losing the feasibility of the outcome. The estimated typical values of $L$, $C$, and $R$ (from Eqs. (18)-(20)) are legitimately applied to overlook the negligibly smaller-order terms in the analytical expressions and continue with the reasonably significant terms. The solution of the QE (Eq. (10)) is expressed in terms of $\omega$ (Eq. (15)) with separate real ($\omega_r$) and imaginary ($\omega_i$) parts. The $\omega_r$ (Eq. (16)) and $\omega_i$ (Eq. (17)) depict the growth or decay and propagation or evanescence of the PSW, respectively. The positive values of both $\omega_i$ and $\omega_r$ denote the PSW forming an instability and its energy propagating towards the bulk plasma exciting IAWs.

The applied linear perturbation also helps in classifying the plasma parameters aiding in both instability and stability of the system [38]. In this analysis, the PSW is found to take form of an instability ($\omega_i > 0$) with the sheath current amplitude growing temporally without saturation. Analysing instabilities in plasma sheath using an $LCR$ circuital model is first of its kind as far as reported in the literature. This $LCR$ model analysis further adds to the novelty of this presented work as a unique piece of elements for the first time of its kind found so far.

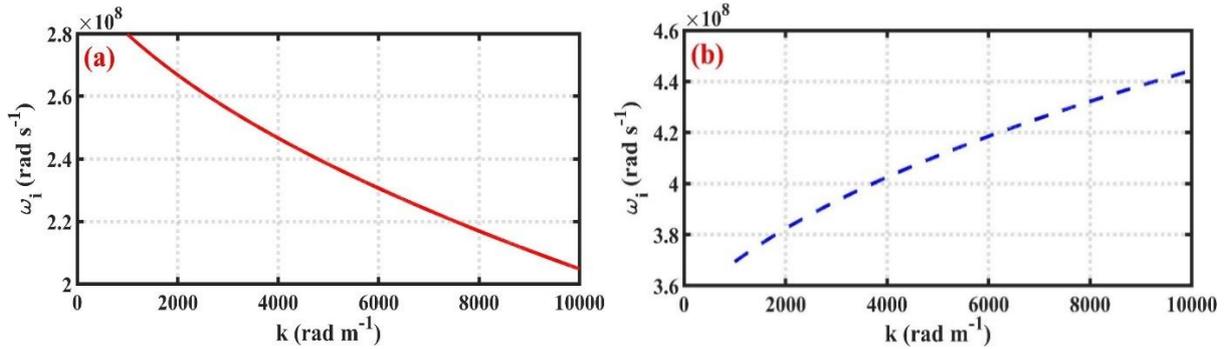

**Figure 2**. Growth rate ($\omega_i$ from Eq. (17)) of the PSW amplitude with the wavenumber ($k$) for two distinct cases: (a) $D > 0$ and (b) $D < 0$.

Regardless of the $D$-polarity chosen (positive or negative, in Eq. (19)), the resultant positive $\omega_i$-value clearly depicts an excited instability in the sheath, expressed in the form of temporally growing PSW amplitude. The variation in $\omega_i$-magnitude points at the variation in the response of the PSWs excited for different $k$-values. The maxima and minima of $\omega_i$ at respective $k$-values insinuate the most and least efficient growth of the instability against the perturbations of different wavenumbers or wavelength. However, irrespective of the chosen polarity and $k$-values, the positive $\omega_i$-magnitude proves the excitation of PSW instability across the sheath.

The absence of any $\omega_i < 0$ in Fig. 2 also points at the sufficiently available free energy in the sheath region for such instabilities. Although $\omega_i > 0$ indicates the instability excitation across the sheath region, modelled via the $LCR$ circuit theory, however, the source of the free energy may vary with the sign (polarity) of imaginary $\omega_c$ discussed in Fig. 3.

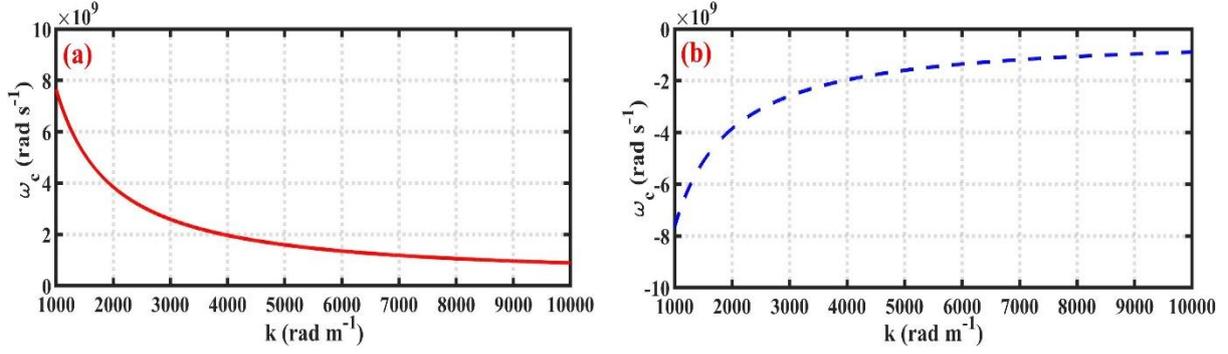

**Figure 3**. Variation of the natural current oscillation frequency ($\omega_c$ from Eq. (34)) through the *LCR* circuit with the wavenumber ($k$) for two distinct cases (a) $\omega_c > 0$ and (b) $\omega_c < 0$.

It must be mentioned herein that the natural current oscillation frequency ($\omega_c$) originates due to the sheath edge oscillation and changing effective *emf* across the sheath. Therefore, the $\omega_c$-nature also manifests the nature of the oscillating sheath or the resultant PSW. The imaginary positive polarity of $\omega_c$ (Eq. (34)) implies an instability excitement inside the sheath region due to the possible energy inflow from the bulk plasma to the sheath. In contrast, an imaginary negative polarity of $\omega_c$ also implies an instability at the sheath region, but via energy outflow from the sheath to the bulk plasma. However, in both cases, an IAW gets excited in the bulk plasma, and it behaves as the carrier of mechanical energy from the sheath to the bulk plasma.

In a nutshell, the imaginary $\omega_c > 0$ implies instability excitation in the sheath region with its self-generated free energy. The available extra free energy is supplied to the bulk plasma through the IAW propagation therein. For imaginary $\omega_c < 0$, it may be inferred that the source of the free energy is not exclusively shrunk within the sheath but a fraction of it also arrives from the bulk plasma through the IAW pulses. However, in both cases ($\omega_c \lessgtr 0$), the instability is analogously getting excited, but with opposite sense of the energy flow patterns.

The resembling magnitudes of $\omega_c$ ($10^9 - 10^{11}$ rad s$^{-1}$) and the IAW angular frequencies prove the energy transfer across these regions through the IAWs. Furthermore, an imaginary $\omega_c$ proves that an introduced small-scale perturbation applied to the plasma sheath yields harmonic spatiotemporal sheath oscillations at the beginning. The oscillating sheath successively undergoes damping with the oscillation energy converted into the IAW in the ambient plasma. The sheath oscillation energy is transferred to the resonating heavier ions in the IAW. The electrons, however, fail to resonate with the oscillating sheath due to their comparatively higher frequency of oscillation. Consequently, the free energy transfer from the sheath to the bulk plasma occurs exclusively through the bulk plasma ions (acting as the inertial species) with the electrons behaving as the corresponding restoring (thermal) species.

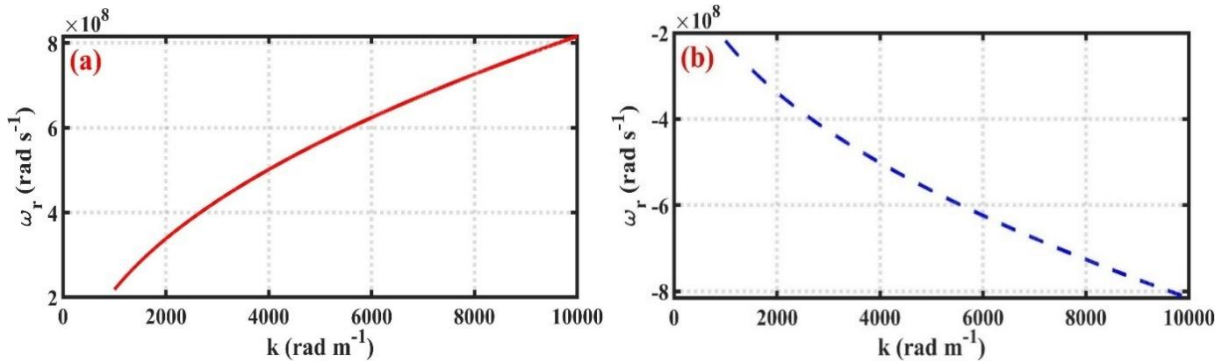

**Figure 4**. Variation of real angular frequency ($\omega_r$ from Eq. (16)) with wavenumber ($k$) for (a) $\omega_r > 0$ (propagatory) and (b) $\omega_r < 0$ (evanescent) nature of the PSW.

The positive and negative polarities of $\omega_r$ in the above figure denote the possible propagative and evanescent nature of the PSW, respectively. A propagative wave enables itself to expel from the source where it excites, whereas an evanescent wave gets diminished within the source itself. In either of the cases the PSW can trigger IAW as the sheath and bulk plasma regions are adjacent to each other. The $\omega_r$-variation (along y-axis) across the wavenumber ($k$) denotes the variable response of the PSW for different values of the wavenumber or wavelength of the perturbations in the medium. The PSW propagates or shrinks itself for some particular $k$-windows in the wave parameter space.

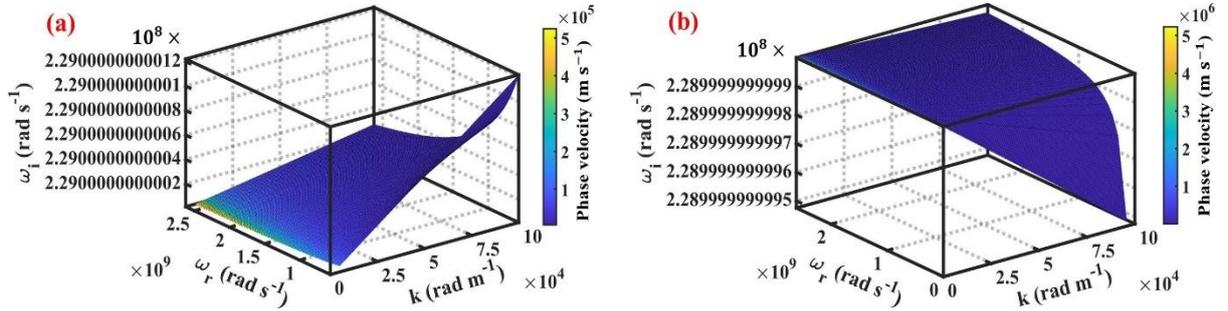

**Figure 5**. Variation of imaginary angular frequency ($\omega_i$ from Eq. (17)) with respect to real angular frequency ($\omega_r$ from Eq. (16)) for different values of wavenumber ($k$). The two different figures correspond to (a) $D > 0$ and (b) $D < 0$. The minimal change along the z-axis at thirteenth place from the decimal denotes a very sensitive inter-dependence of $\omega_i$ on $\omega_r$, and vice-versa.

The colourmap above is generated considering $\omega_r > 0$ to examine the impact of propagative PSW for both positive and negative $D$-values (Eqs. (18)-(19)). A very meagre variation is noticed to exist in the $\omega_i$-magnitude (along z-axis in Fig. 5). This small variation in the $\omega_i$-magnitude with respect to $\omega_r$ (along y-axis) implies the distinct behavior of the PSW in terms of its amplitude growth, decay, and propagation, or evanescence (although relatively small). It also suggests that the propagative or evanescent nature of the PSW does not dominantly influence its instability growth or decay under any circumstance and vice versa. Both $\omega_r$- and $\omega_i$-variational properties vary individually with respect to $k$. However, the color variation in the surface plot and colorbar (on right) displays the phase velocity variation of the PSW relative to $\omega_r$ and $k$. As seen before (Figs. 2, Fig. 4), the 4-D colourmap (Fig. 5) manifests the collective variation of all the derived PSW parameters with respect to the defined wave parameters. The 4 dimensions of the colormaps in Fig. 4 (now, Fig. 5 in the revised version) are $k$, $\omega_r$, $\omega_i$, and $v_p$ (phase velocity as color variation). The fourth dimension is denoted by the color mapping gradient.

      The parametric variations in graphs (Figs. 2-5) and self-illustrating intermediate texts demonstrate the novelty of this circuital model by their resembling results with the ones reported previously. Furthermore, the mutually independent propagative or evanescent and instability nature of the PSW evident from Fig. 5 adds another originality to the *LCR* sheath analysis.

**Conclusions**

The *LCR* circuital model of plasma sheath presented herein examines the behavior of dynamic sheath and successive PSW formations. The application of conventional linear small-scale perturbation formalism yields a QE in terms of angular frequency ($\omega$) from the governing second-order DE. The systematically gained analytic solution of the QE using some valid approximations based upon the realistic parametric values of $L$, $C$, and $R$ yields feasible results. The results corroborate with the established literature published before on PSW characteristics and successful energy transfer to the bulk plasma forming the IAW therein. The fair matching of the natural current oscillation frequency ($\omega_c$) with that of the IAW frequency proves the IAW formation in the bulk plasma originating from the oscillating sheath or the PSW.

In a nutshell, the *LCR* plasma sheath circuital model analysis demonstrates that, due to the sheath width fluctuation, there is a resultant current fluctuation in the system, and vice-versa. The sheath width fluctuation triggers the PSWs due to the spatiotemporal variation of the electric field across it. The excited PSWs manifested in the form of density inhomogeneities and fluctuations in the adjacent sheath region provides energy to the bulk plasma region. The energy gets transported across bulk plasma through the IAW excitation and propagation as already proven experimentally [39]. The PSW may trigger an instability at $\omega_i > 0$ (Fig. (2)).

Furthermore, the number of applications of PSW and sheath-plasma interaction across the fields (environs) proves the applied values of this research. Some applications of controlled sheath oscillations comprise of etching, material deposition, plasma propulsion, and so forth. Besides, sheath oscillation yields several outcomes that may have both fundamental as well as applicational values. Some of these are discussed briefly below.

(i) Ion energy modulation: The time dependent sheath thickness in RF plasma discharge systems (PDSs), such as CCPs, significantly influences the ion energy distribution towards the test substrate [40]. Moreover, in the dual frequency CCPs, the interaction between the low- and high-frequency components can be utilized to modulate ion energies allowing for more precise control over the ion energy distributions [41]. The same experiment has also been conducted with a nonzero phase difference between the two applied driving frequencies. The results show that phase angle enables independent tuning of the ion energy and flux [42,43]. Furthermore, the experiments are also performed by customizing the voltage waveforms applied to the electrodes. It helps in achieving the desired ion-energy distribution (IED) profiles suitable for specific material processing requirements in technological perspective [44,45]. It is observed that an increase in the electrode potential fluctuation frequency to a very high value reduces the maximum ion energy [46-52]. It helps in finding the most efficient biasing frequencies [22]. The circuital model helps in figuring out the necessary conditions in terms of $\omega_c$ for efficient ion energy modulation in such PDSs.

(ii) Electron heating: The electron heating process through plasma discharge systems (PDSs), such as CCPs, has several relevant applications, such as in semiconductor etching [23], thin film deposition [24], surface modification [53], plasma ashing [54], plasma polymerization [55], etc. The standard techniques of heating electrons for respective applications are given below:

(a) Ohmic (collisional) heating: It occurs due to collisions in between electrons and neutral atoms, leading to energy absorption from the RF electric field emanating from the installed electrodes in the PDS [56].

(b) Stochastic (collisionless) heating: It occurs by the interactions of the oscillating sheath electric fields with that of electrons during the rapid expansion and contraction of the sheath in the PDSs [57].

(c) Pressure heating: It occurs due to pressure gradient generated from sheath expansion and contraction, leading to energy transfer to electrons in the PDSs [58].

(d) Transit-time (acoustic) heating: It occurs through electrons traversing regions of varying electric field near the sheath edge in the PDSs [59].

(e) Nonlinear and resonant heating: It occurs through complex interactions between electrons and electric fields, including the generation of higher-order harmonics and subsequent resonance effects [60].

The major factors which influence the electron heating significantly are the driving frequency of electrode electric fields, sheath dynamics, plasma density, and pressure [61-67], etc. The circuital model helps in figuring out the necessary conditions in terms of the ion implantation current frequency ($\omega_c$) for efficient electron heating in PDSs.

(iii) Instability excitation: The sheath oscillations in contact with the bulk plasma can excite diverse plasma waves, such as ion-acoustic, electron acoustic waves, etc. If the sheath oscillation amplitude becomes supercritical (greater than the excitation threshold value), then various instabilities get triggered [38]. For instance, having a nonplanar sheath and a pervading magnetic field may yield inhomogeneous fluid flow. It can trigger the shear-driven Kelvin Helmholtz instability, as observed in the astroplasmic Heliospheric regions. The shear arises due to different velocities of the individual plasma layers [68]. This encourages extrapolating the *LCR* sheath circuit model also to the astroplasmic domain with relevant modifications (if needed).

(iv) Plasma-surface interaction: Sheath oscillations may increase ion bombardment frequency on any introduced surface in the plasma medium, expediting applications like ion-assisted etching or thin-film deposition. Uncontrolled sheath oscillation, however, may cause erosion and damage the material. For instance, it is found that for sheaths of potential ~100 V in high-density plasmas, localized RF power deposition can reach a level of material damage [69]. Thus, learning the plasma-surface interaction and PSW (as done herein) has profound application in this applied field [23,24].

(v) Signal distortion: The sheath oscillation developed around antennas installed in astroplasmic environments (e.g., electrodynamic tethers and space stations) may distort the signal emanating from them. The electromagnetic harmonics generated due to nonlinearities may make the signal fuzzy and perplexed for analysis [70].

(vi) Non-invasive plasma diagnosis: Analysis of electromagnetic properties of the plasma sheath using *LCR*-like models can be helpful in extracting information regarding various plasma parameters without using any kind of conventional plasma probes directly. This analysis helps in understanding wave attenuation and density variation preserving the integrity of the plasma system with better efficiency, which, otherwise, gets compromised with usual sheath-enveloped probes [71]. Moreover, in the case of electrostatic wave-sheath interactions, the number density variation and sheath oscillations are found to be temporally varying and inextricably related to collective plasma analysis [19].

(vii) Higher harmonic generation (HHG): In the CCP systems, the application of sinusoidal RF voltage pulse leads to the formation of oscillating plasma sheaths near the electrodes. The nonlinear dynamics of the sheaths manifested in terms of contraction and expansion can distort the applied RF waveform. This results in a generation of higher harmonics, which are integer multiples of applied driving RFs. These higher harmonics can influence plasma behavior, including electron heating, ion energy distribution, and collective plasma stability [72-81]. A few significant concluding remarks about the HHGs based on the literature are cast below:

  (a) Electric field filamentation and HHG: Investigations on electric field filamentation and HHG in very high-frequency capacitive discharges show that at elevated driving frequencies the electric field becomes spatially non-uniform, thereby leading to filamentation. This non-uniformity enhances the generation of higher harmonics, which, in turn, affects the electron heating mechanisms and plasma uniformity [72].

  (b) Voltage- vs. current-driven discharges: Through comparisons of voltage-driven and current-driven CCPs, it is found that the voltage-driven case exhibits more pronounced HHGs due to the excitation of plasma series resonance (PSR), thereby leading to an enhanced nonlinear electron power dissipation. Whereas the current-driven discharges are noticed to suppress the PSR, resulting in reduced harmonic content. It proves the relevance of the RF-driving methods on HHGs and plasma behaviors [73].

  (c) Electromagnetic effects and HHG: Through the experiments on electromagnetic wave propagations and HHGs in CCPs, it is noticed that electromagnetic effects become significant at higher frequencies and larger electrode areas, leading to the excitation of higher harmonics. These harmonics can interact with the plasma, thereby affecting its uniformity and the efficiency of power deposition [74].

  (d) Driving frequency and HHG: It is observed that increasing the electrode biasing driving frequency enhances the HHGs, which contributes to more effective electron heating and can alter the electron energy distribution function. This relationship emphasizes the role of driving frequency in controlling the HHGs and plasma characteristics [75].

  (e) Nonlinear effects: Nonlinearities introduced by substrate biasing can lead to the generation of higher harmonics, influencing ion energy distributions and potentially impacting the surface processing outcomes [73,76].

The practical implications of HHGs in CCP systems comprise of process control (semiconductor manufacturing), plasma uniformity (surface treatments across substrates), equipment design (designing power delivery systems to mitigate unwanted harmonic effects), and so forth [77-81].

In epilogue, this proposed work is intended to be studied further in the nonlinear domain also, which includes studying the saturation mechanism possible in the system. It is expected that similar sheath-plasma interaction model studies are further to be extended to astroplasmic systems, bearing both basic and practical values on a similar physically applied footing.

**Data availability**
All data generated or analyzed during this study are included in this published article.

**Acknowledgements**
The authors are thankful for fruitful discussions with Dr. Chandra Bhushan Dwivedi, ex-Scientist and ex-Faculty, CPP-IPR, India. They also acknowledge the technical assistance received from fellow researchers of the Astrophysical Plasma and Nonlinear Dynamics Research Laboratory (APNDRL), Department of Physics, Tezpur University, Tezpur, India.

**Author contributions**
Pralay Kumar Karmakar originated, formulated, and supervised the research problem entirely as an active corresponding author. Subham Dutta carried out analytical calculations and numerical analyses as a dynamic contributing author. Both authors contributed significantly to preparing this manuscript in the current refined form.

**DECLARATIONS**

**Competing interests**
The authors declare no competing interests.

**Additional information**
Correspondence and requests for materials should be addressed to Pralay Kumar Karmakar (email: pkk@tezu.ernet.in).

**Appendix**: IAW energy density calculation

It may be interesting to evaluate here the total amount of energy carried by an excited IAW in the plasma chamber with an oscillating sheath. The total energy density of IAW per unit volume ($\varepsilon$) can be calculated by adding the individual ion kinetic energy per unit volume ($m_i n_o < v_i^2 >/2$) and electrostatic energy per unit volume ($\epsilon_o < E^2 >/2$) of the wave with all usual notations cast as

$$\varepsilon = \frac{1}{2} m_i n_o < v_i^2 > + \frac{1}{2} \epsilon_o < E^2 >. \qquad (A(1))$$

Here, $m_i$, $n_o$, $v_i$, $\epsilon_o$, and $<>$ denote ion mass, equilibrium ion number density, ion velocity, permittivity of free space, and time average over one cycle, respectively. Using judicial values of the relevant plasma parameters for an exemplary hydrogen plasma system, we can anticipate a feasible value of energy for a considered volume of space.

To find the term $< E^2 >$, we assume an IAW equation of the following form

$$\phi(r,t) = \phi_o \cos(kr - \omega t), \qquad (A(2))$$

The negative gradient of $\phi$ ($-\partial_r \phi$), yields an electric field, $E$ as follows

$$E = k\phi_o \sin(kr - \omega t). \qquad (A(3))$$

It is clear from Eq. (A(3)) that the time average of $E^2$ is $k^2 \phi_o^2 / 2$. Therefore, we have

$$< E^2 > = k^2 \phi_o^2 / 2. \qquad (A(4))$$

Moreover, we assume the ion velocity, $v_i$ equivalent to the ion sound phase speed, $c_s$ mentioned previously as, $c_s = \sqrt{k_B T_e / m_i}$. Using $v_i \approx \sqrt{k_B T_e / m_i}$ and $< E^2 > = k^2 \phi_o^2 / 2$ from Eq. (A(4)) in Eq. (A(1)), we have the IAW energy density as a sum of the kinetic (thermal) and potential (electrostatic) counterparts cast as

$$\varepsilon = \frac{1}{2} n_o k_B T_e + \frac{1}{4} \epsilon_o k^2 \phi_o^2, \qquad (A(5))$$

Here, $k$ and $\phi_o$ denote the wavenumber and amplitude of the electric potential variation of the IAW, respectively. Using an exemplary $n_o = 10^{16}$ m$^{-3}$, $k_B T_e = 10$ eV, $\epsilon_o = 8.85 \times 10^{-12}$ F m$^{-1}$, $k = 10^4$ rad s$^{-1}$, and $\phi_o = 10$ V, we have

$$\varepsilon \approx 8 \times 10^{-3} + 22.12 \times 10^{-3} \approx 0.03 \text{ J m}^{-3}.$$

Therefore, it enables us to infer that the energy density carried by the IAW with an oscillating sheath in the defined plasma system is given as $\varepsilon = 3 \times 10^{-2}$ J m$^{-3}$.